# Hysteretic ac loss in a coated superconductor subjected to oscillating magnetic field: ferromagnetic effect and frequency dependence

Guang-Tong Ma

*Applied Superconductivity Laboratory, State Key Laboratory of Traction Power, Southwest Jiaotong University, Chengdu, Sichuan 610031, China*

**Abstract**

Numerical simulations of the hysteretic ac loss in a coated superconductor with a more realistic version of architecture were performed via finite-element technique in the presence of an oscillating magnetic field. The coated superconductor was electromagnetically modeled by resorting to the quasistatic approximation of a vector potential approach in conjunction with the nonlinear descriptions of the superconducting layer and ferromagnetic substrate therein by a power law model and the Langevin equation respectively. A diverse effect of the ferromagnetic substrate on the hysteretic ac loss, depending on the strength of the applied magnetic field, was displayed and its underlying cause was identified. The dependence of the hysteretic ac loss on the applied frequency is found to be related to a critical amplitude of the applied magnetic field, and the eddy-current loss dissipated in the metal coatings becomes prominent as the frequency augments merely at high applied magnetic fields.

**PACS numbers**: 84.71.Mn; 74.25.N-; 85.25.Am; 03.50.De

## 1. Introduction

Coated superconductors have been focused for years towards a wide range of applications [1–7] as the persistent advance in the related materials technology [8].One of the most critical properties regarding their use is the hysteretic ac loss caused by exotic stimulations, namely, imposed ac transport currents and/or applied oscillating magnetic fields. As for the topic of applying magnetic fields from a theoretical point of view, a number of studies have been carried out by means of either analytical calculations [9–11] or numerical analysis [10, 12–20], and considerable results in terms of the hysteretic ac loss with ferromagnetic effect or its dependence upon the applied frequency have been achieved as yet. However fundamental these results may be, further investigations at a more realistic level remain promising as a surplus adaption or hypothesis of the geometrical and material characteristics has been made in the existing models, e.g., the thickness of the superconducting layer being scaled due to the intractable aspect ratio of width/thickness of a real coated superconductor [10, 14]; the ferromagnetic substrate being supposed as a linear media with constant or even infinite permeability [9, 14, 17]; the field-dependent feature of the critical current, which is found to be tangible at high applied magnetic fields [21], being neglected or silent in the literatures [9, 12–14, 17]; the superconductor being described by exploiting the magnetostatic–electrostatic analogs in order to adapt to the commercial software ANSYS [17].

Motivated by the situation described above, this paper is dedicated to take a comprehensive examination of the ferromagnetic effect and frequency dependence on the hysteretic ac loss of a coated superconductor, with a pristine geometry and containing all electromagnetically critical constituents of a real one rather than the simplified version, i.e., a superconductor strip over a substrate [9, 12, 13, 15–20], by making use of the quasistatic approximation of a vector potential approach in conjunction with an elaborate description of the nonlinear behaviors in both the superconducting layer and ferromagnetic substrate.

## 2. Mathematical model

Referring to the commercially attainable product from SuperPower Inc. [22], the coated superconductor addressed in this work is supposed to be made up of four constituents to denote a more realistic version of material architecture, viz., superconducting layer as the central element, silver overlayer, metallic substrate and copper stabilizer, all being infinitely extended in the $z$-axis of a Cartesian coordinate system $x$, $y$, $z$, as demonstrated in figure 1. The frequency of the applied magnetic field covered by this work is limited to be in a low range (1–500 Hz), rendering the quasistatic approximation of the Maxwell's equations valid and permitting the omission of the frequency-related characterization of the magnetic permeability and electrical conductivity as well.

Given the premise of this sort, the electromagnetic master equation for the coated superconductor as well as the surrounding coolant is established in terms of the magnetic vector potential **A** by using Ampere's law within the quasistatic approximation as the state equation,

$$\nabla \times \left( \frac{1}{\mu} \nabla \times \mathbf{A} \right) = -\sigma \frac{\partial \mathbf{A}}{\partial t} \tag{1}$$

where the magnetic permeability $\mu$ in the ferromagnetic substrate and the electrical conductivity $\sigma$ in the superconducting layer are nonlinear. The 2-D reduced form of equation (1) in different materials has been presented in details elsewhere [23].

An inverse representation of the power law [24], combined with Kim's model [25], is adopted to characterize the nonlinear dependence of the supercurrent density on the local fields,

$$\mathbf{J} = \frac{J_{c0}}{1+|\mathbf{B}|/B_0} \left( \frac{|\mathbf{E}|}{E_c} \right)^{1/n} \frac{\mathbf{J}}{|\mathbf{J}|}, \tag{2}$$

where $J_{c0}$ is the zero-field critical current density depending on the prescribed criterion $E_c$, and $n$ is the creep exponent, whereas $B_0$ represents a critical magnetic flux density for which the critical current density is halved. Understanding the superconducting layer is made of yttrium-barium cuprate cooled with liquid nitrogen, the present analysis has used $J_{c0} = 2.5 \times 10^{10}$ A/m$^2$ or $I_{c0} = 100$ A [22], $E_c = 1$ $\mu$V/cm, $n = 21$ [26], and $B_0 = 0.1$ T [27]. A residual resistivity $\rho_0 = 10^{-17}$ $\Omega$m, to account for the flux creep due to the thermal activation in the superconductor as well as to insure the numerical stability around the zero electric field [28, 29], is assigned to the superconducting layer in this work.

If the ferromagnetic substrate is made of Ni-based alloy, whose *B-H* characteristic shows a minor loop that permits the neglect of coercivity [30, 31], calling upon a reversible-paramagnet approximation in the Langevin form [32],

$$|\mathbf{B}| = \mu_0 \left\{ M_s \left[ \coth\left( \frac{|\mathbf{H}|}{H_0} \right) - \frac{H_0}{|\mathbf{H}|} \right] + |\mathbf{H}| \right\}, \tag{3}$$

with the saturation magnetization $M_s$ and the auxiliary magnetic field strength $H_0$ linked to the magnetic susceptibility at zero field $\chi_0$ by $H_0 = M_s/3\chi_0$, is logically suggested. The hysteretic loss in the ferromagnetic substrate is therefore not taken into account here to be self-consistent with such hypothesis, though an empirical formula for that is already attainable [15, 19]. It is presumed that $M_s = 7.5 \times 10^5$ A/m and $\chi_0 = 250$ [23, 33] in this paper for a ferromagnetic substrate, unless stated otherwise.

The hysteretic ac loss per unit length in a full cycle of applied magnetic field is computed as,

$$U_{ac} = \oint_{f.c.} dt \iint_{\Omega} \mathbf{E} \cdot \mathbf{J} \, dxdy, \qquad (4)$$

where $\Omega$ is the cross-sectional area of the respective domain of each constituent in the coated superconductor.

The electromagnetic master equation (1) is numerically discretized by means of the Galerkin's finite-element method [34] and Euler's finite-difference scheme [35], respectively, in the spatial and temporal domain, and the generated nonlinear system of finite-element equation is solved by resorting to the Jacobian-free Newton-Krylov algorithm, an advanced approach founded on a synergistic combination of Newton-type methods for superlinearly convergent solutions of nonlinear equations and Krylov subspace methods for solving the Newton correction equations [36]. The oscillating magnetic field, with amplitude $\mu_0 H_a$ and frequency $\upsilon$, is imposed by attaching a Dirichlet condition on the outer bounds of the whole computational domain, a square having a side length of 20 times width of the coated superconductor.

## 3. Results and discussion

With the above-described theoretical foundations, numerical simulations of appraising the effect of the ferromagnetic substrate on the hysteretic ac loss of a coated superconductor subjected to an oscillating magnetic field, together with its frequency dependence, were carried out by assigning the representative geometrical and material characteristics aforementioned to the coated superconductor of figure 1. These simulations also include the eddy-current loss dissipated in the metal coatings using the suggested cryogenic resistivity of pure Ag (0.27 $\mu\Omega$ cm), Ni (0.5 $\mu\Omega$ cm) and Cu (0.19 $\mu\Omega$ cm) to respectively represent the silver overlayer, the metallic substrate and the copper stabilizers [37].

*3.1. Ferromagnetic effect*

The hysteretic ac loss of a coated superconductor with varied ferromagnetic substrate, plus a reference with nonmagnetic substrate, were calculated as a function of the amplitude of an applied transverse magnetic field with $v = 50$ Hz and the normalized results in the electromagnetic steady state (all established since the second cycle) were plotted in figure 2. Through observing this figure, whereas the general trait of whichever property of the substrate is uniform, the effect of the ferromagnetic substrate on the normalized hysteretic ac loss $U_{ac}/(\mu_0 H_a)^2$, suffered by the entire coated superconductor, is evident and exhibits three distinct characters depending on the amplitude of applied magnetic field $\mu_0 H_a$. For small values of $\mu_0 H_a$, the ferromagnetic effect brings out an increase of the hysteretic ac loss in comparison with the reference, being particularly pronounced as the ferromagnetic property of the substrate, controlled by the value of $M_s$, strengthened and the value of $\mu_0 H_a$ abates. Conversely, the hysteretic ac loss for moderate values of $\mu_0 H_a$ is suppressed due to the ferromagnetic effect, firstly building up and then trailing off as the value of $\mu_0 H_a$ augments. Eventually, the hysteretic ac loss of all cases asymptotically converges for large values of $\mu_0 H_a$, the ferromagnetic effect becoming insignificant because of the magnetization saturation of a practical ferromagnet. The set of curves as an inset, representing the variation of the normalized hysteretic ac loss in the superconducting portion only, behaves similarly in both tendency and magnitude for small and moderate values of $\mu_0 H_a$ as those of the entire coated superconductor, implying that (i) the ferromagnetic effect mostly acts on the superconducting layer and (ii) the eddy-current loss dissipated in the coatings only become tangible for large values of $\mu_0 H_a$, providing the geometrical and material characteristics in this work.

It is worth noting that, the increase of the hysteretic ac loss, for small values of $\mu_0 H_a$ in the presence of a ferromagnetic substrate, is likely to be attributed to the edge effect due to the magnetic concentration of ferromagnet as a slight extension of width of substrate, by merely a factor of 1.1 to shape a wider substrate, will lead to a substantial decrease of the hysteretic ac loss, being completely below that of nonmagnetic case, as clearly demonstrated in figure 2. However, widening the substrate will accordingly enhance the portion of eddy-current loss due to the increase of the conducting volume and as a result, cause the hysteretic ac loss of the entire coated superconductor be slightly higher than others for large values of $\mu_0 H_a$, as figure 2 displays. This unfavorable effect can be dramatically mitigated as the cryogenic resistivity of substrate augments, which has been proven by using a higher resistivity, for Ni–5%W substrate at 77 K [38]. These findings, revealed on a theoretical version of coated superconductor with all critical constituents present and relying on the finite-element technique with power law current–voltage model and nonlinear permeability, alongside the previous predictions of a superconductor strip on a ferromagnetic substrate using an analytic model with critical state model and infinite permeability [9], perhaps suggest that, the ferromagnetic effect is capable of reducing the hysteretic ac loss in a coated superconductor exposed to oscillating magnetic field, at least for the small and moderate values of $\mu_0 H_a$, by carefully tuning the width of substrate. (An adjunctive calculation of the scenario addressed by the analytical model [9] and experiments [39], putting the present investigations into an established perspective, was also carried through and the related results can be found in the Appendix)

*3.2. Frequency dependence*

Figure 3 portrays the normalized hysteretic ac loss $U_{ac}/(\mu_0 H_a)^2$, suffered by the entire coated superconductor and the superconducting layer therein in the electromagnetic steady state (also established since the second cycle), against the amplitude of an applied transverse oscillating magnetic field $\mu_0 H_a$, addressing a series of values of frequency $\upsilon$ within a limited band (1–500 Hz). A general feature revealed by this figure, irrespective of the uniform tendency in terms of the amplitude $\mu_0 H_a$ for whatever value of the frequency $\upsilon$, is that, there exists a critical value of $\mu_0 H_a$, over and below which the dependence of the hysteretic ac loss on the frequency is quite distinct. A growing increase with the frequency of the hysteretic ac loss in the entire coated superconductor is seen to occur as the value of $\mu_0 H_a$ augments from the critical point, whereas a decreasing trend emerges for the values of $\mu_0 H_a$ below the critical point, excluding the exceptions at small values of $\mu_0 H_a$, where the dependence becomes irregular. In contrast, the hysteretic ac loss in the superconducting portion, demonstrated as an inset in figure 3, displays a regular variation with increasing the frequency, being respectively elevated and degraded at the values of $\mu_0 H_a$ over and below the critical point. Providing the geometrical and material characteristics of the coated superconductor chosen in this work, the threshold value of $\mu_0 H_a$, distinguishing the distinct variations of the hysteretic ac loss with the amplitude of the applied magnetic field, is estimated to be slightly less than 25 mT for the case of that in the entire coated superconductor, while for the case of that in the superconducting layer, is to be slightly higher than 25 mT, according to the numerical interval for $\mu_0 H_a$ in the present simulations.

The extracted data from figure 3 to clearly display the variation of the hysteretic ac loss with frequency at a certain value of $\mu_0 H_a$ was sketched and presented in figure 4 for that in the superconducting layer (left), in the copper stabilizers (middle) and in the entire coated superconductor (right). The hysteretic ac loss as a function of frequency, $U_{ac}(\upsilon)$, was normalized by that of $\upsilon_0 = 1$ Hz in each case of this figure. The value of $U_{ac}(\upsilon_0)$ for normalization is therefore different among the three graphs in figure 4. The left figure reveals that, the above-described decrease and increase of the hysteretic ac loss in the superconducting layer with the frequency, respectively below and over the critical point, both develop as an exponential dependence that becomes more pronounced while the value of $\mu_0 H_a$ approaching the extremes, whereas at the value of $\mu_0 H_a$ around the critical point, being slightly higher than 25 mT, the variation of the frequency only gives rise to a tiny change in the hysteretic ac loss. The middle figure shows that, the eddy-current loss in the copper stabilizers increases linearly with the frequency, but the slope is distorted as compared to the theoretically expected value of one [38], being weakly at the extremes of $\mu_0H_a$ but rather markedly at the intermediate values of $\mu_0H_a$. These findings, to some extent, are consistent with the previous studies [10]. The dependence of the total loss in the entire coated superconductor, illustrated in the right figure, indicates that, the hysteretic ac loss in the superconducting layer is dominant at the small values of $\mu_0H_a$, where an exponential decrease of the total loss with the frequency is displayed, whereas at the large values of $\mu_0H_a$, a quasilinear increase in the hysteretic ac loss emerges, implying that the contribution of the eddy-current loss in the metal coatings becomes prominent.

Figure 5 shows the dependence of the normalized hysteretic ac loss $U_{ac}/(\mu_0H_a)^2$ in the respective metal coating on the amplitude of an applied transverse magnetic field $\mu_0H_a$ to demonstrate how the screening effect [40] due to the induced supercurrent in the superconducting layer affects these dependences at different frequencies. It can be seen from this figure that, given any value of the frequency, the screening effect mostly acts at small and moderate values of $\mu_0H_a$, particularly towards the silver for which a roughly linear increase is found, whereas at large values of $\mu_0H_a$, the normalized hysteretic ac loss nearly keeps constant as it should [38], implying an insignificant screening effect.

Worthy of comment is that, though all results above presented were obtained by applying a transverse oscillating magnetic field, the main observations were estimated to be held with other field orientations, only the peak of the curves in figures. 2 and 3 being shifted as those released in reference [18], according to the achieved results for other orientations of the applied magnetic field.

## 4. Conclusion

In conclusion, the ferromagnetic effect on the hysteretic ac loss of a coated superconductor subjected to oscillating magnetic field, together with the frequency-dependent features, has been examined by means of a numerical model in this paper. It is mainly observed that, the ferromagnetic substrate mostly affects the superconducting layer and its effect on the hysteretic ac loss is diverse depending on the strength of the applied magnetic field, a phenomenon recalling the measurement at hand [39, 41, 42]. An inverse dependence of the hysteretic ac loss on the applied frequency exists over and below a critical amplitude of the applied magnetic field, and as the applied frequency augments, the eddy-current loss dissipated in the metal coatings tends to be prominent merely at high applied magnetic fields. The screening effect, due to the induced supercurrent in the superconducting layer, on the metal coatings is found to be significant only at the small and moderate applied magnetic fields, irrespective of the applied frequency.

## Acknowledgements

This work was supported in part by the National Natural Science Foundation of China under Grant 51007076, and by the State Key Laboratory of Traction Power at Southwest Jiaotong University under Grant 2013TPL_T05, and by the Alexander von Humboldt Foundation.

## Appendix. Superconductor strip on an idealized ferromagnetic substrate

Considering a bilayer heterostructure of a thin superconductor strip sitting on a ferromagnetic substrate and quoting the characteristics released in references [9] and [39], the hysteretic ac loss inside the superconductor strip was calculated as a function of the amplitude of an applied transverse oscillating magnetic field with frequency $\upsilon = 20$ Hz relying on the finite-element technique with a nonlinear current-field dependence ($n = 21$ and $B_0 \rightarrow \infty$) and a high constant magnetic permeability ($\mu_r = 10^4$), and the normalized results of the imaginary part of ac susceptibility $\chi''$, estimated by $\chi'' = U_{ac} / \pi \mu_0 H_0^2$ [9] with $\mu_0 H_0$ representing the amplitude of the applied magnetic field, were plotted in figure A.1. Shown also in this figure is that of a nonmagnetic substrate and a widened ferromagnetic substrate to create the scenario addressed in reference [9]. The eddy-current loss in the substrate is neglected in these calculations to be in line with the analytical counterparts [9].

It is seen clearly from figure A.1 that the results, achieved by the present finite-element analysis, well reproduce the main features demonstrated in figure 5(b) of reference [9], particularly the excellent consistency in terms of the intersection between the curves for the superconductor strip sitting on a nonmagnetic substrate and a ferromagnetic substrate with identical width to the strip, occurring at $\mu_0 H_0 \cong 0.48$ mT or $H_0/j_c d_s \cong 0.138$ by the present estimation versus $\mu_0 H_0 \cong 0.49$ mT or $H_0/j_c d_s \cong 0.14$ in the previous prediction [9], which may be considered as evidence to verify the present finite-element analysis and, on the other hand, to support the validity of the analytical model proposed in reference [9]. (The symbols used in reference [9] is cited in the Appendix for direct comparison)

**Figures**

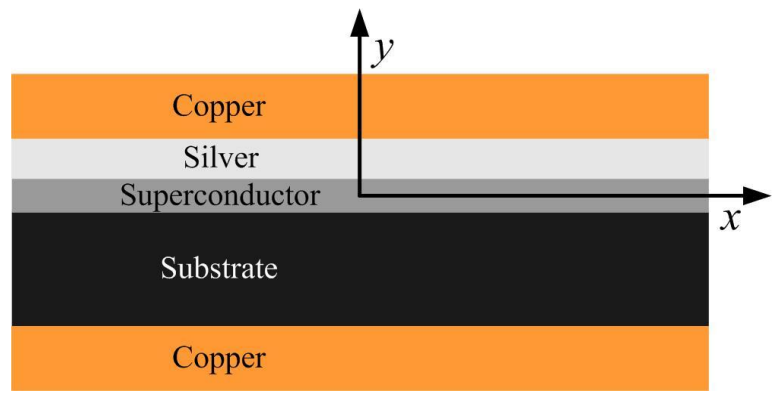

**Figure 1.** Cross-sectional view of an infinitely extended coated superconductor with the superconducting layer covered by a silver cap and deposited upon a metallic substrate and then sandwiched by top and bottom copper stabilizers. The thickness of each constituent from top to bottom is respectively 20, 2, 1, 50, 20 $\mu$m, with an identical width of 4 mm, referring to the SCS405 conductor from SuperPower Inc. [22]. The dimensions shown in this drawing are not to scale.

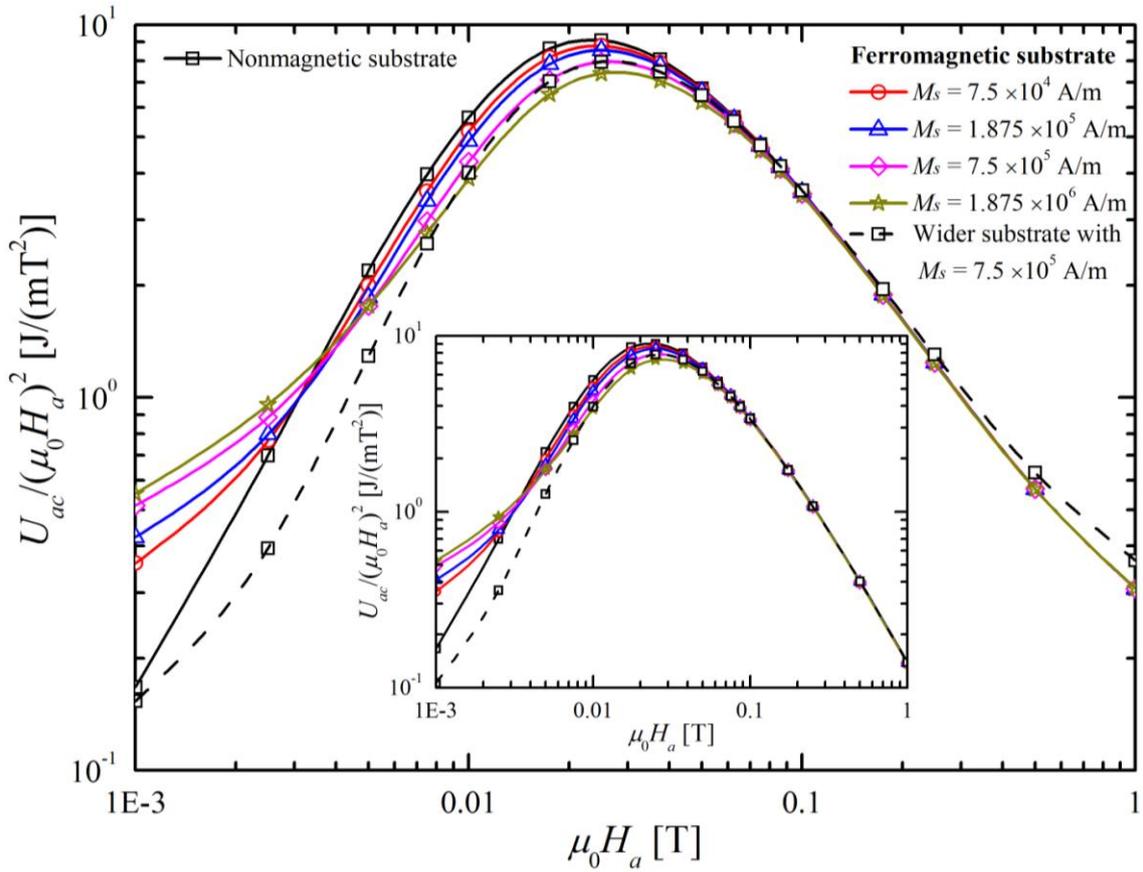

**Figure 2.** Normalized hysteretic ac loss, suffered by the entire coated superconductor and by the superconducting layer (inset) per cycle, as a function of the amplitude of applied transverse oscillating magnetic field with $v = 50$ Hz in the electromagnetic steady state (all established since the second cycle). Different ferromagnetic properties were assigned to the substrate through tuning the saturation magnetization $M_s$. Also shown for comparison is that by a coated superconductor with nonmagnetic substrate (solid-rectangle curve) and by a coated superconductor with wider substrate (dashed-rectangle curve). The wider substrate has a dimension in width of 4.4 mm, being 1.1 times of the original one shown in figure 1.

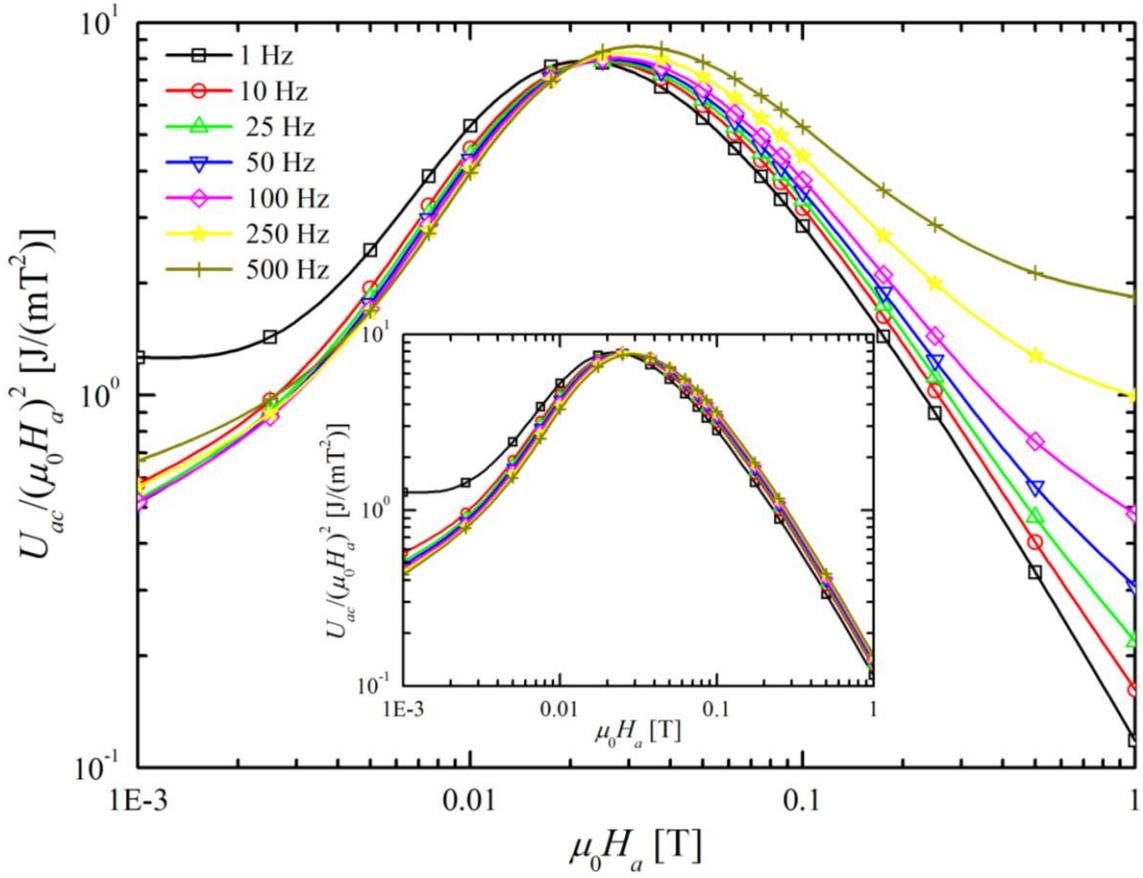

**Figure 3.** Normalized hysteretic ac loss, suffered by the entire coated superconductor and by the superconducting layer (inset) per cycle, as a function of the amplitude of applied transverse oscillating magnetic field with varied frequency (1–500 Hz) in the electromagnetic steady state (all established since the second cycle).

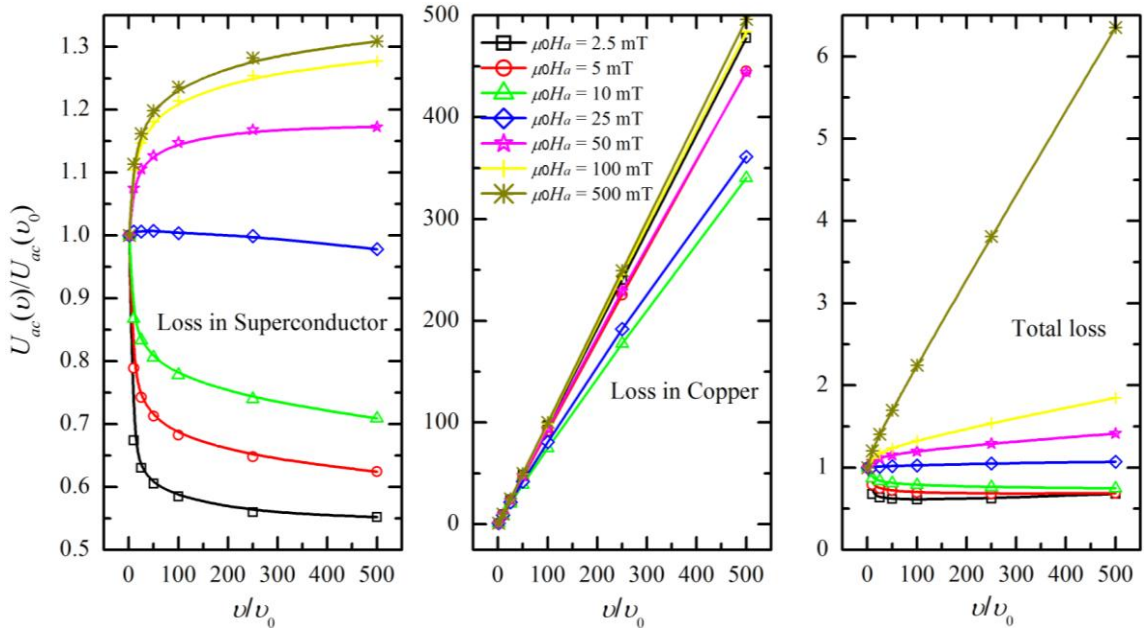

**Figure 4.** Normalized hysteretic ac loss, suffered by the superconducting layer (left), the copper stabilizers (middle), and the entire coated superconductor (right) per cycle, as a function of the normalized frequency of applied transverse oscillating magnetic field with different amplitudes $\mu_0 H_0$ in the electromagnetic steady state (all established since the second cycle). The value of frequency for normalization $v_0$ is 1 Hz and the value used for normalization, $U_{ac}(v_0)$, is different among the three graphs.

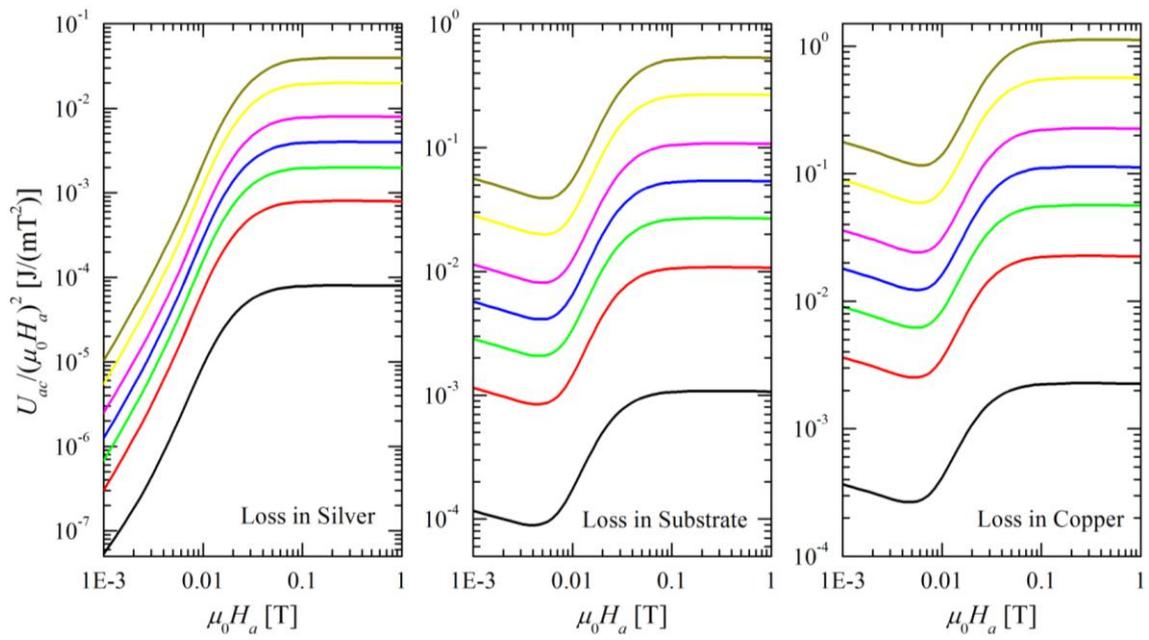

**Figure 5.** Normalized hysteretic ac loss, suffered by the silver overlayer (left), the ferromagnetic substrate (middle), and the copper stabilizers (right) per cycle, as a function of the amplitude of applied transverse oscillating magnetic field with $v$ = 1, 10, 25, 50, 100, 250, 500 Hz (from the lower curve up) in the electromagnetic steady state (all established since the second cycle).

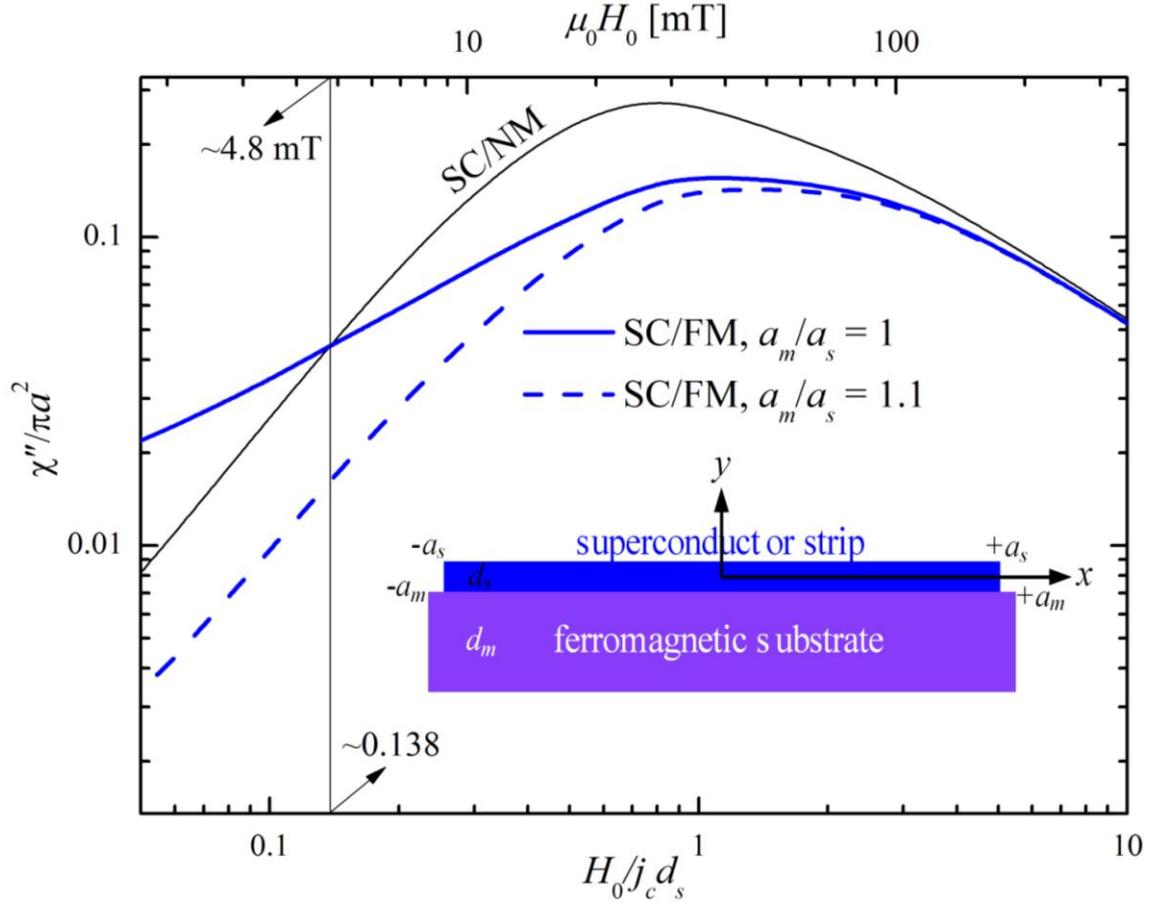

**Figure A.1.** Normalized imaginary part of ac susceptibility $\chi''/\pi a^2$ as a function of the amplitude of an applied transverse oscillating magnetic field $\mu_0 H_0$, or the related normalized value $H_0/j_c d_s$, for a superconductor (SC) strip sitting on a nonmagnetic (NM) substrate (thin solid line), or on a ferromagnetic (FM) substrate with $a_m/a_s = 1$ (thick solid line), or with $a_m/a_s = 1.1$ (dashed line) in the electromagnetic steady state (established since the second cycle). The dimensions of $a = a_s = 5$ mm, $d_s = 2.3$ μm, and $d_m = 25$ μm, together with the critical current density of $j_c = 1.2 \times 10^{10}$ A/m$^2$, as those in references [9] and [39], were adopted in the present calculations. The relative permeability of ferromagnetic substrate here was assumed to be constant and as high as $10^4$ to approach the infinite assumption made in reference [9].